\documentclass[preprint,showpacs,amsmath,amssymb]{revtex4}
\usepackage{graphicx}
\usepackage{dcolumn}
\usepackage{bm}
\usepackage{psfig}
\newcommand{\ssb}{\Sigma^0\overline{\Sigma^0}}
\newcommand{\aab}{\Lambda\overline{\Lambda}}
\newcommand{\gaa}{\gamma\Lambda\overline{\Lambda}}
\newcommand{\BR}{{\cal B}}
\newcommand{\eff}{\varepsilon}

\newcommand{\probpi}{Prob_{\pi}}
\newcommand{\probka}{Prob_{K}}
\newcommand{\probpr}{Prob_{p}}

\newcommand{\psp}{\psi(2S)}
\newcommand{\jpsi}{J/\psi}
\newcommand{\chicJ}{\chi_{cJ}}
\newcommand{\chicz}{\chi_{c0}}
\newcommand{\chico}{\chi_{c1}}
\newcommand{\chict}{\chi_{c2}}
\newcommand{\psipp}{\pi^+\pi^- J/\psi}

\newcommand{\pip}{\pi^+}
\newcommand{\pim}{\pi^-}
\newcommand{\piz}{\pi^0}
\newcommand{\pp}{\pi^+\pi^-}
\newcommand{\ppb}{p\overline{p}}

\newcommand{\pppr}{\pi^+\pi^-p\overline{p}}

\newcommand{\gpppr}{\gamma\pi^+\pi^-p\overline{p}}

\newcommand{\jpsipp}{\pi^+\pi^-J/\psi}
\newcommand{\ppjpsi}{\pi^+\pi^-J/\psi}
\newcommand{\ra}{\rightarrow}
\newcommand{\jpsito}{J/\psi \rightarrow }

\newcommand{\pspto}{\psi(2S) \rightarrow }
\newcommand{\chicJto}{\chi_{cJ} \rightarrow }
\newcommand{\chiczto}{\chi_{c0} \rightarrow }

\newcommand{\chictto}{\chi_{c2} \rightarrow }
\newcommand{\bfg}{\begin{figure}}
\newcommand{\efg}{\end{figure}}
\newcommand{\bitm}{\begin{itemize}}
\newcommand{\eitm}{\end{itemize}}
\newcommand{\bnum}{\begin{enumerate}}
\newcommand{\enum}{\end{enumerate}}
\newcommand{\btbl}{\begin{table}}
\newcommand{\etbl}{\end{table}}
\newcommand{\btbu}{\begin{tabular}}
\newcommand{\etbu}{\end{tabular}}
\begin{document}

\preprint{Draft-PRD}

\title{\boldmath First Evidence of $\aab$ in $\chicJ$ Decays}
\author{
J.~Z.~Bai$^1$,        Y.~Ban$^{8}$,          J.~G.~Bian$^1$,
X.~Cai$^{1}$,          J.~F.~Chang$^1$,
H.~F.~Chen$^{14}$,    H.~S.~Chen$^1$,
J.~Chen$^{7}$,        J.~C.~Chen$^1$,     
Y.~B.~Chen$^1$
% ...
\\(BES Collaboration)\\ 
$^1$ Institute of High Energy Physics, Beijing 100039, People's Republic of
     China\\
$^2$ China Center of Advanced Science and Technology, Beijing 100080,
     People's Republic of China\\
$^3$ Henan Normal University, Xinxiang 453002, People's Republic of China\\
$^4$ Huazhong Normal University, Wuhan 430079, People's Republic of China\\
$^5$ Hunan University, Changsha 410082, People's Republic of China\\
$^6$ Liaoning University, Shenyang 110036, People's Republic of China\\
$^7$ Nankai University, Tianjin 300071, People's Republic of China\\
$^{8}$ Peking University, Beijing 100871, People's Republic of China\\
$^{9}$ Shandong University, Jinan 250100, People's Republic of China\\
$^{10}$ Shanghai Jiaotong University, Shanghai 200030, 
        People's Republic of China\\
$^{11}$ Sichuan University, Chengdu 610064,
        People's Republic of China\\       
$^{12}$ Tsinghua University, Beijing 100084, 
        People's Republic of China\\
$^{13}$ University of Hawaii, Honolulu, Hawaii 96822\\                       
$^{14}$ University of Science and Technology of China, Hefei 230026,
        People's Republic of China\\
$^{15}$ Wuhan University, Wuhan 430072, People's Republic of China\\
$^{16}$ Zhejiang University, Hangzhou 310028, People's Republic of China\\
\vspace{0.4cm}
$^{\ast}$ Visiting professor to University of Michigan, Ann Arbor, MI 48109 USA 
}

\date{Apr. 7, 2003}
           
\begin{abstract}
  The first observation of $\chicJ$ (J=0,1,2) decays to $\aab$ is
  reported using $\psp$ data collected with the BESII detector at the
  BEPC. The branching ratios are determined to be \( \BR(\chicz
  \rightarrow \aab) = (4.7^{+1.3}_{-1.2} \pm 1.0)\times 10^{-4}\) , \(
  \BR(\chico \rightarrow \aab) = (2.6^{+1.0}_{-0.9} \pm 0.6)\times
  10^{-4}\) and \( \BR(\chict \rightarrow \aab) = (3.3^{+1.5}_{-1.3}
  \pm 0.7)\times 10^{-4}\).  Results are compared with model
  predictions.
\end{abstract}

\maketitle


\section{Introduction}

It has been shown both in theoretical calculations and experimental
measurements that the lowest Fock state expansion (color singlet
mechanism, CSM) of charmonium states is insufficient to describe
P-wave quarkonium decays. Instead, the next higher Fock state (color
octet mechanism, COM) plays an important role~\cite{so,width}.  Our
earlier measurement~\cite{width} of the total width of the $\chicz$ agrees
rather well with the COM expectation.  The calculation of the partial
width of $\chicJto \ppb$, by taking into account the COM of $\chicJ$
decays and using a carefully constructed nucleon wave
function~\cite{wong}, obtains results in reasonable agreement with
measurements~\cite{pdg}.  The nucleon wave function was then
generalized to other baryons, and the partial widths of many other
baryon anti-baryon pairs predicted. Among these predictions, the
partial width of $\chicJto \aab$ is about half of that of $\chicJto
\ppb$~(J=1,2)~\cite{wong}.

In this paper, we report on an analysis of the $\gpppr$ final state
produced in $\psp$ decays. Evidence for the decays of $\chicJ$ to $\aab$ is
observed for the first time.
The data used for this analysis were taken with the Beijing
Spectrometer detector (BESII) at the Beijing Electron Positron
Collider (BEPC) at a center-of-mass (CM) energy corresponding
to $M_{\psp}$. The data sample corresponds to a total of about 15
million $\psp$ decays.

BES is a conventional solenoidal magnet
detector that is described in detail in Ref.~\cite{bes}; BESII is the
upgraded version of the BES detector~\cite{bes2}. A 12-layer vertex
chamber (VTC) surrounding the beam pipe provides trigger information.
A forty-layer main drift chamber (MDC), located radially outside the
VTC, provides trajectory and energy loss ($dE/dx$) information for
charged tracks over $85\%$ of the total solid angle with a momentum
resolution of $\sigma _p/p = 0.0178 \sqrt{1+p^2}$ ($p$ in $\hbox{\rm
GeV}/c$) and a $dE/dx$ resolution for hadron tracks of $\sim 8\%$.
An array of 48 scintillation counters surrounding the MDC measures the
time-of-flight (TOF) of charged tracks with a resolution of $\sim 200$
ps for hadrons.  Radially outside the TOF system is a 12 radiation
length, lead-gas barrel shower counter (BSC).  This measures the
energies of electrons and photons over $\sim 80\%$ of the total solid
angle with an energy resolution of $\sigma_E/E=21\%/\sqrt{E}$ ($E$ in
GeV).  Outside the solenoidal coil, which provides a 0.4~Tesla
magnetic field over the tracking volume, is an iron flux return that
is instrumented with three double layers of counters that identify
muons of momentum greater than 0.5~GeV/c.

A Monte Carlo simulation is used for the determination of the mass
resolution and detection efficiency, as well as the estimation of the
background.  For the signal channels, $\pspto \gamma \chicJ$,
$\chicJto \aab$, the angular distribution of the photon emitted in the
$\psp$ decay is assumed to be that for a pure E1 transition.  The
$\Lambda$ in the $\chicJ$ CM system and the daughter particles in the
$\Lambda$ CM system are generated isotropically. A total of 10000
events are generated for each $\chicJ$ state with $\Lambda
\rightarrow \pi^- p$ and $\overline{\Lambda}\rightarrow \pi^+
\overline{p}$.
For the estimation of the number of $\psp$ events and the estimation
of the systematic error, $\pspto \jpsipp$, $\jpsito \ppb$ events are
generated, where the $\pi^+\pi^-$ invariant mass is distributed as
measured in Ref.~\cite{bll}.

The simulation of the detector response, including interactions of
secondary particles in the detector material, uses a
Geant3 based package SIMBES. Reasonable agreement between
data and Monte Carlo simulation is observed in testing various
channels, including $e^+ e^- \rightarrow \gamma e^+ e^-$ (Bhabha),
$e^+ e^- \rightarrow \mu^+ \mu^-$, $\jpsito \ppb$, and $\pspto
\jpsipp$, $\jpsito \ell^+\ell^-$.

\section{Event selection}

The analysis uses the same photon selection and charged particle
identification (ID) criteria as were used in Ref.~\cite{had}.
When selecting photons it is necessary to remove photons 
produced by hadronic interactions of charged tracks
with the detector material.  This is achieved by cutting on the 
angle between the neutral cluster and the charged track in the BSC.
The number of photon candidates in an event is not limited.

Both TOF and $dE/dx$ information are used for charged particle
identification.  Probabilities of a track being a pion ($\probpi$),
kaon ($\probka$), or proton ($\probpr$) are assigned to each charged
track.
For the decay channel of interest, the candidate events are 
required to satisfy the following selection criteria:
\begin{enumerate}
\item   Each charged track is required to be well fit to a
        three-dimensional helix and be in the polar angle region 
        $|\cos\theta_{MDC}|<0.8$.
\item   The number of charged tracks is four with net charge zero. 
\item   The two lower momentum positive and negative charged tracks are
        assumed to be the $\pi^+$ and the $\pi^-$, and the other two tracks
        are regarded as the proton and the antiproton. Four-constraint
        kinematic fits to the decay hypothesis are performed with each of
        the photon candidates, and the one with the smallest $\chi^2$ is
        taken as the real photon.  The $\chi^2$ probability of the fit is
        required to be greater than 1\%.
      \item The particle identification assignment of each charged
        track must satisfy $\probpi$ (for $\pi^{\pm}$) or $\probpr$
        (for $p$ or $\overline{p}$) $> 0.01$.
\end{enumerate}

A four-constraint fit assuming $\pspto \pppr$ is also performed 
to select $\pspto \aab$ and $\pspto \ppjpsi, \jpsito \ppb$ events for 
checking the reliability of the analysis of $\chicJto \aab$ and
to calculate the total number of $\psp$ events.
The selection criteria used are the same as for $\chicJto \aab$ except
that no photon information is used.

\section{Event analysis}

Fig.~\ref{ma-ma} shows a scatter plot of the $\pi^+\overline{p}$
versus the $\pi^-p$ invariant mass for events with $\pppr$ mass
between $3.38$~GeV/$c^2$ and $3.60$~GeV/$c^2$. The cluster of events
in the lower left corner shows a clear $\aab$ signal.

\begin{figure}[htbp]
\centerline{\hbox{
\psfig{file=ma-ma.epsi,width=10.00cm}}}
\caption{Scatter plot of $\pi^+\overline{p}$ versus $\pi^-p$ invariant
mass for selected $\gpppr$ events with the $\pppr$ mass in the $\chicJ$ mass 
region.}
\label{ma-ma}
\end{figure}

\begin{figure}[htbp]
\centerline{\hbox{
\psfig{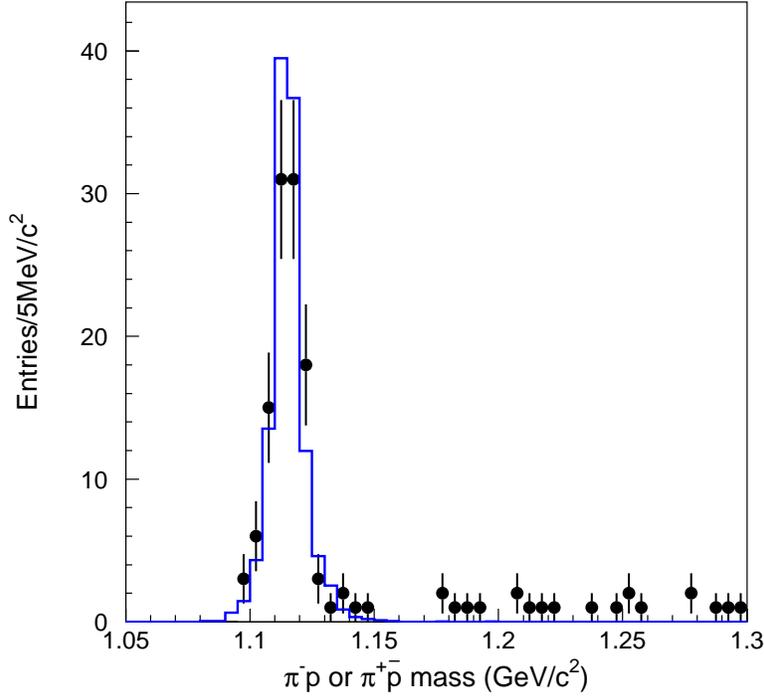}}}
\caption{Mass distribution of $\pi^+\overline{p}$ ($\pi^-p$)
recoiling against a $\Lambda$ ($\overline{\Lambda}$)
(mass $<1.15$~GeV) for events in the $\chicJ$ mass region.
Dots with error bars are data and the histogram is the
Monte Carlo simulation, normalized to the $\Lambda$ signal
region (two entries per event).}
\label{ma}
\end{figure}

Selecting events in $\chicJ$ mass region and requiring the mass
of $\pi^+\overline{p}$ ($\pi^-p$) to be smaller than 1.15~GeV/$c^2$, the
$\pi^-p$ ($\pi^+\overline{p}$) mass distribution shown in
Fig.~\ref{ma} is obtained. A clear $\Lambda$ signal can be seen, and the 
background below the peak is very small.  A fit of the mass
distribution gives $m_{\Lambda}= (1114.6 \pm 0.6)~\hbox{MeV}/c^2$,
in agreement with the world average~\cite{pdg},
and a mass resolution of $(6.3\pm 0.6)~\hbox{MeV}/c^2$.

\begin{figure}[htbp]
\centerline{\hbox{
\psfig{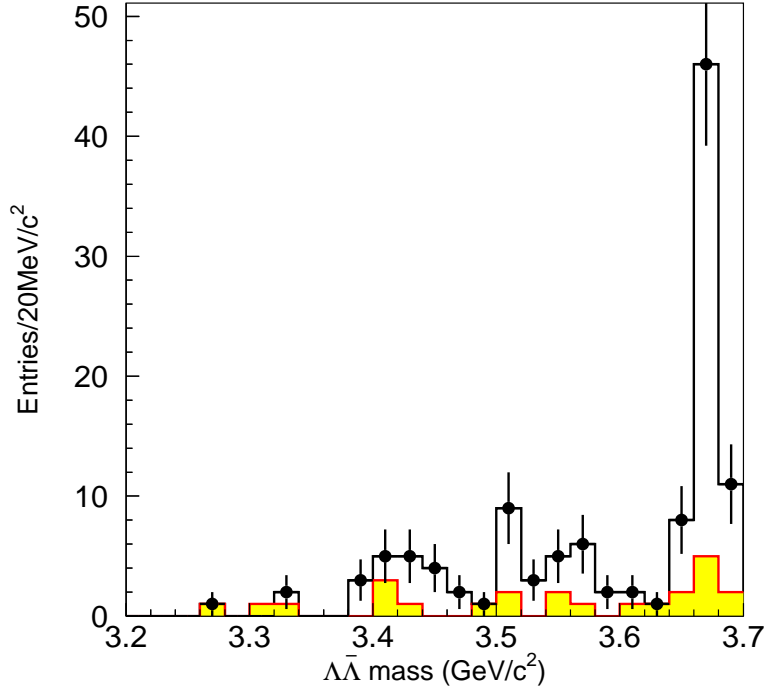}}}
\caption{Mass distribution of $\aab$ candidates. Histogram with error
bars is data, and the shaded histogram is from $\Lambda$ side bands 
events (normalized).}
\label{maaside}
\end{figure}

After requiring that both the $\pi^+\overline{p}$ and the $\pi^-p$ mass
lie within twice the mass resolution around the nominal $\Lambda$
mass, the $\aab$ invariant mass distribution shown in
Fig.~\ref{maaside} is obtained. There are clear $\chicz$, $\chico$, and
$\chictto \aab$ signals with low background, estimated using $\Lambda$
mass side band events. The highest peak around the $\psp$ mass is due
to $\pspto \aab$ with a fake photon.

Fig.~\ref{sce} shows the energy deposited in the BSC of the proton
track versus the antiproton track for events selected as $\chicJto
\aab$.  Since the antiproton will frequently annihilate in the detector,
much of the energy of the annihilation products may be
detected in the BSC. The scatterplot is consistent with these
expectations, indicating the two tracks are really the proton and
anti-proton.

\begin{figure}[htbp]
\centerline{\hbox{
\psfig{file=sce.epsi,width=10.00cm}}}
\caption{Energy deposited in the BSC of the proton and
antiproton tracks for selected $\chicJto \aab$ events.}
\label{sce}
\end{figure}

\begin{figure}[htbp]
\centerline{\hbox{
\psfig{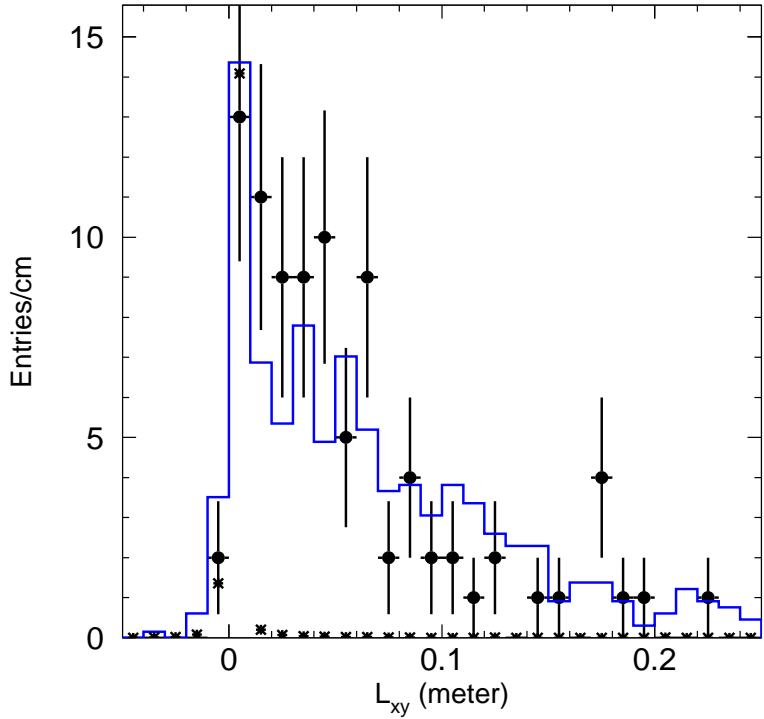}}}
\caption{Secondary vertex distributions. Dots with error bars 
  are for events with their mass in the $\chicJ$ mass region, the
  histogram is for selected $\pspto \aab$ events, and the asterisks
  are for selected $\pspto \jpsipp$, $\jpsito \ppb$ events.  The dots
  and histogram are normalized for greater than 1 cm, and the
  normalization for the asterisks is arbitrary. }
\label{lxy}
\end{figure}

Fig.~\ref{lxy} shows the distribution of secondary vertices in the
$xy$ plane of $\gaa$ candidates in the $\chicJ$ mass region (error
bars).  This distribution shows good agreement with the secondary
vertex distribution of selected $\pspto \aab$ events (histogram), but
is significantly different from the vertex distribution of $\pspto
\jpsipp$, $\jpsito \ppb$ events (stars), where no secondary vertex is
expected. This indicates the events in the $\chicJ$ mass region are real
$\aab$.

\subsection{Remaining backgrounds}
\begin{figure}[htbp]
\centerline{\hbox{
\psfig{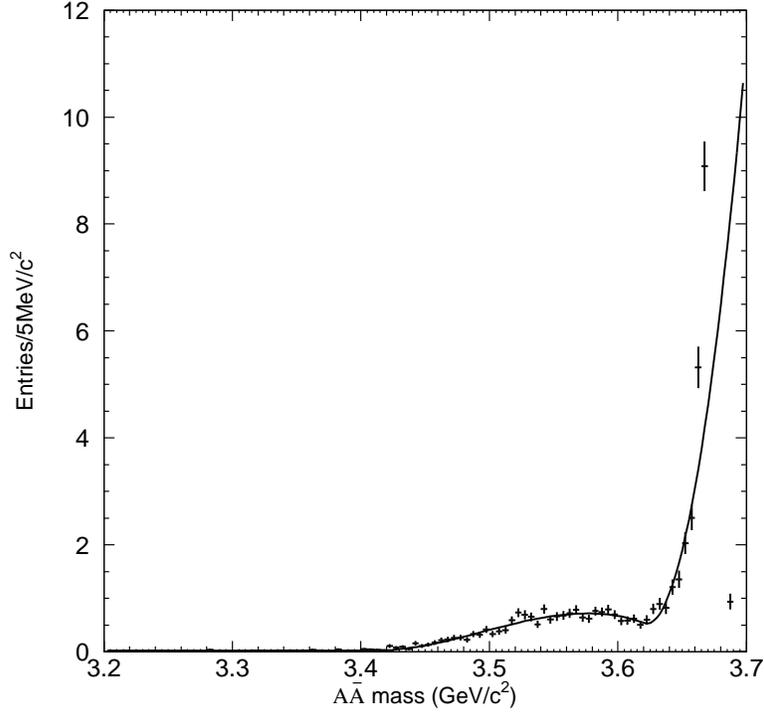}}}
\caption{Invariant mass distribution of $\aab$ selected from
Monte Carlo simulated background events normalized to the total number of
$\psp$ events in the data sample. The curve shows the best fit
of the mass spectrum below 3.65~GeV/$c^2$.}
\label{bkg_shape}
\end{figure}

Background from non $\aab$ events is 
estimated from the $\Lambda$ mass sidebands as shown in
Fig.~\ref{maaside}, and this can be described in fitting the 
$\aab$ mass spectrum by a linear background.
The background from channels with $\aab$ production, 
including $\pspto \aab$, $\pspto \ssb$, 
$\pspto \Lambda \overline{\Sigma^0} + c.c.$,
$\pspto \Xi^0 \overline{\Sigma^0} + c.c.$, $\pspto \gamma \chicJ,
\chicJto \Sigma^0\overline{\Sigma^0} \ra \gamma \gamma \aab$, and
$\pspto \psipp \rightarrow \pppr$, are simulated by Monte Carlo. 
By using the branching ratios of $\pspto \aab$, $\pspto \ssb$, and
$\pspto \psipp \rightarrow \pppr$ measured by previous 
experiments~\cite{pdg}, and naively assuming 
$\pspto \Lambda \overline{\Sigma^0} + c.c.$ and 
$\pspto \Xi^0 \overline{\Sigma^0} + c.c.$ are one order of 
magnitude smaller than $\pspto \aab$ and $\pspto \ssb$, and
$\chicJto \Sigma^0\overline{\Sigma^0}$ is about the same as
$\chicJto \aab$, we obtain the expected total background plotted in 
Fig.~\ref{bkg_shape}. The curve 
in this plot indicates the best fit of the background mass spectrum from 
3.2 to 3.65~GeV/$c^2$.
The background from events with more
photons is smaller, and Monte Carlo simulation of 
$\pspto \Xi^0 \overline{\Xi^0}$ indicates that its contamination to the 
$\chi_c$ signal is negligible.

\subsection{Fit to the mass spectrum}
\begin{figure}[htbp]
\centerline{\hbox{
\psfig{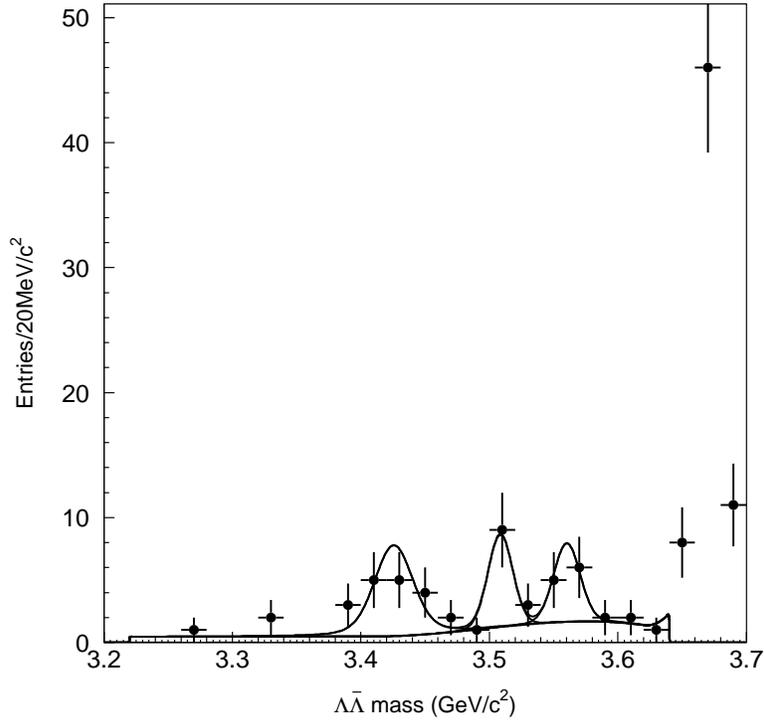}}}
\caption{Mass distribution of $\gaa$ candidates fitted with three
resolution smeared Breit-Wigner functions and background, as
described in the text.}
\label{maafit}
\end{figure}

Fixing the mass resolutions at their Monte Carlo predicted values ($(12.7\pm
0.9)~\hbox{MeV}/c^2$, $(9.4\pm 0.3)~\hbox{MeV}/c^2$ and $(9.8\pm
0.4)~\hbox{MeV}/c^2$ for $\chicz$, $\chico$ and $\chict$,
respectively), and fixing the widths of the three $\chicJ$ states to
their world average values~\cite{pdg}, the mass spectrum was fit with
three Breit-Wigner functions folded with Gaussian resolutions and
background, including a linear term representing the non $\aab$
background and a component described in the previous subsection
representing the $\aab$ background with the global normalization
factor floating to take into account possible systematic bias in the
background estimation (mainly branching ratio uncertainties).  The
unbinned maximum likelihood method was used to fit the events with
$\aab$ mass between 3.22 and 3.64~$\hbox{GeV}/c^2$, and a likelihood
probability of 27\% was obtained, indicating a reliable fit. The
number of events with errors determined from the fit are
$15.2^{+4.2}_{-4.0}$, $9.0^{+3.5}_{-3.1}$, and $8.3^{+3.7}_{-3.4}$ for
$\chicz$, $\chico$ and $\chict$, respectively. The statistical
significances of the three states are $4.5\sigma$, $3.5\sigma$ and
$2.6\sigma$.  Fig.~\ref{maafit} shows the fit result, and the fitted
masses are $(3425.6\pm 6.3)\hbox{MeV}/c^2$, $(3508.5\pm
3.9)~\hbox{MeV}/c^2$ and $(3560.3\pm 4.6)~\hbox{MeV}/c^2$ for
$\chicz$, $\chico$ and $\chict$, respectively, in agreement with the
world average values~\cite{pdg}. The detection efficiencies from the
Monte Carlo simulation were determined to be $\eff_{\chicz}^{MC} =
(6.07 \pm 0.24)\%$, $\eff_{\chico}^{MC} = (6.65 \pm 0.25)\%$ and
$\eff_{\chict}^{MC} = (6.09 \pm 0.24)\%$, where the errors come from
the limited statistics of the Monte Carlo samples. 

\section{\boldmath Number of $\psp$ events}

The number of $\psp$ events is determined using 
$\pspto \ppjpsi, \jpsito \ppb$. There are many advantages in
using this channel to determine the number of events:
\bnum
\item It has the same kind of charged tracks as the channel
      of interest, and the momenta in these 
      two channels are similar, so that in the branching 
      ratio measurement, the systematic bias in tracking, 
      kinematic fit, triggering, particle ID, geometric acceptance of
      charged tracks, etc. will cancel out.
\item It is easy to select, and the error on the branching ratio
      is small ($(2.12\pm 0.10)\times 10^{-3}$ for the
      world average)~\cite{pdg}.
\enum

The selection criteria of this channel are the same as for the 
$\chicJto \aab$ analysis, except the photon is not considered.
The invariant masses of $\pim p$ and $\pip \overline{p}$ are required 
to not be in the $\Lambda$ mass region to remove $\pspto \aab$
background. Fig.~\ref{mjpsi} shows the $\ppb$ invariant mass
distributions of both
data and Monte Carlo.  There is a huge $\jpsi$ signal on top of
very low background.

\begin{figure}[htbp]
\centerline{\hbox{
\psfig{file=mjpsi.epsi,width=10.00cm}}}
\caption{Distribution of $\ppb$ invariant mass of $\pspto \ppjpsi,
\jpsito \ppb$ data (top) and Monte Carlo (bottom).}
\label{mjpsi}
\end{figure}

The number of $\jpsito \ppb$ events is estimated by subtracting
sideband events for $\ppb$ invariant mass regions from 3.0 to 3.05~GeV/c$^2$
and from 3.15 to 3.2~GeV/c$^2$ from the signal region 
($\ppb$ invariant mass from 3.05 to 3.15~GeV/c$^2$), giving
     \[ n^{obs}_{\jpsito \ppb} = 1826 \pm 44. \]
Using the same method, the efficiency is determined using Monte Carlo data
as
     \[ \eff = (17.88\pm 0.12)\%. \]
     Using the BES branching ratio for $\pspto \ppjpsi$ (($32.3\pm
     1.4)\%$~\cite{bespsp}) and the PDG branching ratio for $\jpsito
     \ppb$ ($(2.12\pm 0.10)\times 10^{-3}$~\cite{pdg}), the number of
     $\psp$ events is obtained
\begin{eqnarray}
     N_{\psp} & = & \frac{n^{obs}_{\jpsito \ppb}/\eff}
               {\BR(\pspto \ppjpsi) \BR(\jpsito \ppb)} \nonumber \\
             & = & (14.91\pm 0.36\pm 1.13)\times 10^6, \nonumber
\end{eqnarray}
where the first error is statistical and the second is systematic,
including the statistical error of the efficiency, the errors from the
two branching ratios used, and the uncertainty due to the Monte Carlo
simulation of the angular distributions.

It should be noted that the efficiency correction factors due to the
differences between data and Monte Carlo data in the particle ID, the
kinematic fit, tracking, etc. are not considered, because the same
differences exist in the $\chicJto \aab$ analysis and will cancel in
the $\chicJto \aab$ branching ratio measurement.

As a consistency check, one can apply the particle ID correction factor
($1.043\pm 0.011$) and kinematic fitting correction factor 
($0.943\pm 0.010$), which are measured in following sections. One then
obtains \( N_{\psp} = (15.16\pm 0.37\pm 1.16)\times 10^6 \), which agrees
with the number of $\psp$ events determined using either inclusive
$\pspto \ppjpsi$ or inclusive hadrons.

\section{Efficiency correction and systematic errors}

The systematic errors in the branching ratio measurements come from
the efficiencies of the photon ID, particle ID, kinematic fitting,
low energy photon detection, MDC tracking, the branching
ratios used, the number of $\psp$ events, the $\Lambda$ mass cut, etc.

\subsection{Photon ID}

The fake photon multiplicity distributions in both data and Monte Carlo 
simulation are checked with $\pspto \ppjpsi, \jpsito \ppb$ events. The
Monte Carlo predicts too many fake photons at very low energy 
(less than 50~MeV). Using a photon energy cut at 50~MeV 
or reweighting the Monte Carlo events with the measured fake photon
multiplicity distribution indicates that the Monte Carlo simulates
the data with a precision of 4\%. This will be taken as the systematic 
error on the photon ID.

\subsection{Particle ID}

Samples of $\pip$, $\pim$, $p$, and $\overline{p}$ tracks are selected in
$\pspto \ppjpsi, \jpsito \ppb$ events by requiring a good kinematic fit
to this process and good particle
identification of the other three charged tracks involved.  This
allows a measurement of the particle ID efficiency, and  
a correction factor of $1.043\pm 0.011$ to the Monte Carlo efficiency
is found for the channels that we are studying. The error is from the 
limited statistics of the samples used
and is taken as the systematic error of the particle ID.

\subsection{Kinematic fit}

The bias due to the kinematic fitting is caused by differences between
data and Monte Carlo data in the fitted momentum and error matrix
of the charged
track and differences in
the measurement of the energy and the direction of the neutral track
and their uncertainties. The effect is studied for charged tracks and
neutral tracks separately.

\subsubsection{Charged tracks}

The bias from the 
kinematic fit of the charged tracks was checked 
using $\pspto \ppjpsi$, $\jpsito \ppb$ events. This channel is very
clean and can
be selected without the help of a kinematic fit.
By comparing the number of events with and without a kinematic fit,
the efficiencies for $prob_{\chi^2}>1\%$ are measured to be
$(85.14\pm 0.92)\%$ and $(90.32\pm 0.24)\%$ for data and Monte Carlo, 
respectively. This results in a correction factor for the
Monte Carlo efficiency of $0.943\pm 0.010$ for this specific 
channel.

\subsubsection{Neutral tracks}

The effect of neutral track measurement is studied using
$\pspto \gamma \chicJ, \chicJto \pppr$ events.  A careful
calibration of the neutral cluster information in the
BSC (including the energy and direction measurement
and their errors) was performed using radiative Bhabha 
events from the same $\psp$ data set. By applying this 
calibration to both data and Monte Carlo, the relative 
changes in the branching
ratios of $\chicJto \pppr$ are measured to be
1.1\%, 1.9\% and 4.2\% for $\chicz$, $\chico$ and $\chict$,
respectively. No corrections to the efficiencies are made;
the largest difference (4.2\%) is taken as the systematic 
error in the measurement of neutral tracks.

\subsection{Photon detection efficiency}

The low energy photon detection efficiency is studied with 
$\pspto \ppjpsi$, $\jpsito \pp\piz$ events produced in the same
data sample used for the $\chicJ$ analysis. 
We assume the lower momentum positive and negative charged 
tracks are the $\pi^+$ and $\pim$ from $\psp$ decays, and the 
largest energy neutral cluster is a photon 
from the $\piz$ decay. Assuming the second photon from the $\piz$ decays
is missing, we do a two constraint kinematic fit requiring all
the final particles come from $\psp$ decays and the two photons 
form a $\piz$. The fitted four-momentum of the second
photon is taken as a test beam into the detector and used to determine
the detection efficiency. A total of 2901 photons are selected 
for the efficiency study. The same analysis is performed 
with Monte Carlo events, and agreement between data and Monte Carlo data 
is observed at a precision of $8\%$ for the photons accompanying 
$\chicz$, $\chico$ and $\chict$.

For converted photons, no specific study was performed since this
occurs 
for only a very small fraction of the events (less than 1\%), and the
difference between data and Monte Carlo simulation should be 
even smaller and negligible compared to the quoted systematic error
for the photon efficiencies.

\subsection{Other systematic errors}

The angular distributions of the photon accompanying the $\chicJ$s and
the angular distributions of the $\Lambda$ or $\overline{\Lambda}$
decays may cause a systematic error at the 10\% level.  This is
determined by comparing different theoretical models for the angular
distributions.  The uncertainty in the angular distribution of the
proton in $\jpsi$ decays results in a 4\% error in the determination
of the number of $\psp$ events.

The Monte Carlo simulated mass resolution may have a bias at the 10\%
level. This is determined from the comparison of $\Lambda$ and $\jpsi$
signals in various channels involved in this analysis. Changing the
mass resolutions used in fitting the $\chicJ$ mass plot produces small
changes in the number of events; the maximum change in the three cases
is around 3\%. This is taken as the systematic error due to the mass
resolution uncertainty.

The background estimation, including the uncertainties in the branching
ratios used, the uncertainties in the simulation of the contamination
probability, the parameterization of the background shape, and the 
fitting range used, etc., causes an uncertainty at the 10\% level.
The systematic errors on the branching ratios used, like 
$\BR(\pspto \ppjpsi)$, $\BR(\jpsito \ppb)$, $\BR(\pspto \gamma \chicJ)$ 
and $\BR(\Lambda \ra \pi^- p)$ are obtained from other
experiments~\cite{bespsp,pdg}.

\subsection{Total systematic error}

Table.~\ref{sys} lists the systematic errors from all sources, as well
as the correction factors to the Monte Carlo efficiency for particle
ID and the kinematic fitting of charged tracks.  Since these two
correction factors cancel out in the calculation of branching ratios,
there are no corrections to the efficiencies determined by Monte
Carlo simulation for the $\chicz$, $\chico$ and $\chict$ branching
ratios, and their errors are not considered in the summation.

{\normalsize 
\begin{table}[htbp]
\caption{Summary of systematic errors and the efficiency correction
  factors.   Efficiency correction
  factors are only determined for the particle ID and the kinematic
  fitting of charged
  tracks.  Since these correction factors cancel in the branching
  ratio calculation, they are not used.}
\begin{center}
\begin{tabular}{l|ccc}
\hline\hline
Source  & $\chicz$  &  $\chico$  &   $\chict$  \\\hline
MC statistics &  4.0\%          &  3.8\%          &   4.0\%          \\
Fake photon   & \multicolumn{3}{c}{4\%}  \\
Particle ID       & \multicolumn{3}{c}{1.043$\pm$0.011}  \\
4C-fit (chrg) & \multicolumn{3}{c}{0.943$\pm$0.010}  \\
4C-fit (neut) & \multicolumn{3}{c}{4.2\%}            \\
Phot. eff.    & \multicolumn{3}{c}{8\%}              \\  
Gamma conversion &   \multicolumn{3}{c}{$<$1\%}       \\
Angular distr.   &   \multicolumn{3}{c}{10\%}         \\
Mass resolution  &   \multicolumn{3}{c}{3\%}         \\
Background       &   \multicolumn{3}{c}{10\%}         \\
$\psp$ number    &  \multicolumn{3}{c}{8.0\%}        \\
$\BR(\Lambda \ra \pi^- p)$ & \multicolumn{3}{c}{1.6\%}    \\
$\BR(\pspto \gamma \chicJ)$ & 9.2\%   &  8.3\%   &       8.8\%    \\
\hline
Total systematic error &  22\%  &      21\%      &  22\%         \\\hline
\hline
\end{tabular}
\end{center}
\label{sys}
\end{table}}

\section{Results and discussions}

The branching ratios of $\chicJto \aab$ can be calculated with
\[ \BR(\chicJto \aab)=\left.\frac{n^{obs}/\eff}
     {N_{\psp}\BR(\pspto \gamma \chicJ)\BR(\Lambda\ra \pim p)^2}
     \right. ~. \]
Using numbers from above, one gets
\[ \BR(\chicz \rightarrow \aab)
     = (4.7^{+1.3}_{-1.2} \pm 1.0)\times 10^{-4} ,\]
\[ \BR(\chico \rightarrow \aab)
     = (2.6^{+1.0}_{-0.9} \pm 0.6)\times 10^{-4} ,\]
\[ \BR(\chict \rightarrow \aab)
     = (3.3^{+1.5}_{-1.3} \pm 0.7)\times 10^{-4} ,\]
where the first errors are statistical and the second are 
systematic.  The numbers used and results are summarized in Table.~\ref{br}.

{\normalsize
\begin{table}[htbp]
\caption{Summary of numbers used in the branching ratio calculation and
  branching ratio results. $R_{\BR}$, defined in the text, is the relative branching ratio of
  $\chiczto \aab$ to that of $\pspto \psipp$. }
\begin{center}
\begin{tabular}{l|ccc}
\hline\hline
quantity      & $\chicz$  &  $\chico$  &   $\chict$ \\\hline
$n^{obs}$     & $15.2^{+4.2}_{-4.0}$  
                          & $9.0^{+3.5}_{-3.1}$ 
                                       & $8.3^{+3.7}_{-3.4}$ \\
$\eff$ (\%)   & $6.07 \pm 0.24$ 
                          & $6.65 \pm 0.25$ 
                                       & $6.09 \pm 0.24$ \\
$N_{\psp} (10^6)$           &  \multicolumn{3}{c}{$14.9\pm 1.2$} \\
$\BR(\Lambda \ra \pi^- p)$~\cite{pdg}  &  \multicolumn{3}{c}{$0.639\pm 0.005$} \\
$\BR(\pspto \gamma \chicJ)$ (\%)~\cite{pdg} & $8.7\pm 0.8$
                       & $8.4\pm 0.7$& $6.8\pm 0.6$   \\\hline
$\BR(\chicJto \aab) (10^{-4})$  
              & $4.7^{+1.3}_{-1.2} \pm 1.0$
                         & $2.6^{+1.0}_{-0.9} \pm 0.6$
                                    & $3.3^{+1.5}_{-1.3} \pm 0.7$ \\
\hline\hline
$n^{obs}_{\psipp}$      &  \multicolumn{3}{c}{$1826 \pm 44$} \\
$\eff_{\psipp}$ (\%)     &  \multicolumn{3}{c}{$17.88\pm 0.12$} \\\hline
$R_{\BR}(10^{-2})$  & $2.45^{+0.68}_{-0.65} \pm 0.46$
                         & $1.33^{+0.52}_{-0.46} \pm 0.25$
                                    & $1.33^{+0.59}_{-0.55} \pm 0.25$ \\
\hline
\end{tabular}
\end{center}
\label{br}
\end{table}}

Compared with the corresponding branching ratios of $\chicJto
\ppb$~\cite{pdg}, the branching ratios of 
$\chico$ and $\chictto \aab$ agree with the corresponding 
branching ratios to $\ppb$ within two sigma. This is somewhat 
in contradiction with the expectations from Ref.~\cite{wong},
although the errors are large.

As for $\chiczto \aab$, the measured value agrees with the $\ppb$
measurements from BES and E835~\cite{width,e835} within 2 standard
deviations. One should also note that there is no prediction for
$\BR(\chiczto \aab)$.

What we actually measure in this analysis is the relative branching ratio
of $\chiczto \aab$ to $\pspto \psipp$. 
The
relative branching ratio is found with the following formula
\begin{eqnarray}
R_{\BR} & = &\frac{\BR(\pspto \gamma \chicJ)\cdot 
             \BR(\chicJto \aab)\cdot \BR(\Lambda\ra \pim p)^2}
{\BR(\pspto \psipp) \cdot \BR(\jpsito \ppb)} \nonumber \\
& = &\frac{n^{obs}/\eff}{n^{obs}_{\psipp}/\eff_{\psipp}}. \nonumber
\end{eqnarray}
These results are also shown in Table~\ref{br}.

\section{Summary}

$\aab$ events are observed for the first time in $\chicJ$ decays using
the BESII 15 million $\psp$ event sample, and corresponding branching ratios 
are determined. The results on $\chico$ and $\chict$ decays 
only agree marginally with model predictions.

\acknowledgments

   The BES collaboration thanks the staff of the BEPC for their hard efforts.
This work is supported in part by the National Natural Science Foundation
of China under contracts Nos. 19991480, 10225524, 10225525, the Chinese 
Academy of Sciences under contract No. KJ 95T-03, the 100 Talents Program 
of CAS
under Contract Nos. U-24, U-25, and the Knowledge Innovation Project of 
CAS under Contract Nos. U-602, U-34 (IHEP); by the National Natural Science 
Foundation of China under Contract No.10175060(USTC); and
by the Department of Energy under Contract No DE-FG03-94ER40833 (U Hawaii).

\end{document}